\documentclass[10pt,aps,pra,twocolumn,superscriptaddress]{revtex4-2}
\usepackage{amsmath,amssymb,amsfonts,bm}
\usepackage[breaklinks=true,colorlinks,allcolors=blue]{hyperref}
\usepackage{graphicx}
\usepackage{txfonts}
\usepackage{braket}
\usepackage{comment}
\usepackage{siunitx}
\usepackage{enumitem}
\usepackage{mathtools}
\usepackage[normalem]{ulem}
\usepackage{xcolor}
\usepackage{tabularx, booktabs, multirow}
\usepackage[table]{xcolor}
\usepackage{orcidlink}

\usepackage{booktabs}
\usepackage[ruled,vlined,linesnumbered]{algorithm2e}

\newcommand{\eqlabel}[1]{Eq.~\eqref{#1}}

\newcommand{\tablabel}[1]{Table~\ref{#1}}

\makeatletter
\def\bibsection{%
   \par
   \begingroup
    \baselineskip26\p@
    \bib@device{\hsize}{72\p@}%
   \endgroup
   \nobreak\@nobreaktrue
   \addvspace{19\p@}%
  }%
\makeatother

\begin{document}
\title{Scaling advantage with quantum-enhanced memetic tabu search for LABS}

\date{\today}

\author{Alejandro Gomez Cadavid$^{\orcidlink{0000-0003-3271-4684}}$}
\affiliation{Kipu Quantum GmbH, Greifswalderstrasse 212, 10405 Berlin, Germany}
\affiliation{Department of Physical Chemistry, University of the Basque Country EHU, Apartado 644, 48080 Bilbao, Spain}

\author{Pranav Chandarana$^{\orcidlink{0000-0002-3890-1862}}$}
\affiliation{Kipu Quantum GmbH, Greifswalderstrasse 212, 10405 Berlin, Germany}
\affiliation{Department of Physical Chemistry, University of the Basque Country EHU, Apartado 644, 48080 Bilbao, Spain}

\author{Sebastián V. Romero$^{\orcidlink{0000-0002-4675-4452}}$}
\affiliation{Kipu Quantum GmbH, Greifswalderstrasse 212, 10405 Berlin, Germany}
\affiliation{Department of Physical Chemistry, University of the Basque Country EHU, Apartado 644, 48080 Bilbao, Spain}

\author{Jan Trautmann}
\affiliation{Kipu Quantum GmbH, Greifswalderstrasse 212, 10405 Berlin, Germany}

\author{Enrique Solano$^{\orcidlink{0000-0002-8602-1181}}$}
\email{enr.solano@gmail.com}
\affiliation{Kipu Quantum GmbH, Greifswalderstrasse 212, 10405 Berlin, Germany}

\author{Taylor Lee Patti}
\affiliation{NVIDIA, Santa Clara, California 95051, USA}

\author{Narendra N. Hegade$^{\orcidlink{0000-0002-9673-2833}}$}
\email{narendra.hegade@kipu-quantum.com}
\affiliation{Kipu Quantum GmbH, Greifswalderstrasse 212, 10405 Berlin, Germany}

\begin{abstract}
We introduce quantum-enhanced memetic tabu search (QE-MTS), a non-variational hybrid algorithm that achieves state-of-the-art scaling for the low-autocorrelation binary sequence (LABS) problem. By seeding the classical MTS with high-quality initial states from digitized counterdiabatic quantum optimization (DCQO), our method suppresses the empirical time-to-solution scaling to \(\mathcal{O}(1.24^N)\) for sequence length \(N \in [27,37]\). This scaling surpasses the best-known classical heuristic \(\mathcal{O}(1.34^N)\) and improves upon the \(\mathcal{O}(1.46^N)\) of the quantum approximate optimization algorithm, achieving superior performance with a $6\times$ reduction in circuit depth. A two-stage bootstrap analysis confirms the scaling advantage and projects a crossover point at $N \gtrsim 47$, beyond which QE-MTS outperforms its classical counterpart. These results provide evidence that quantum enhancement can directly improve the scaling of classical optimization algorithms for the paradigmatic LABS problem.
\end{abstract}

\maketitle
\section{Introduction}
Finding low-autocorrelation binary sequences (LABS) is a long-standing and computationally demanding problem at the interface of combinatorial optimization and signal processing~\cite{boehmer1967binary,schroeder1970synthesis}. For a given length, the goal is to find a binary sequence that minimizes the sum of squared off-peak autocorrelations. This objective directly translates into practical performance gains in technologies such as radar, where low sidelobes are crucial for distinguishing targets from background noise. A defining property that makes LABS a stringent benchmark is that there exists exactly one non-trivial instance per sequence length. This removes instance-selection bias and requires algorithms to confront the same increasingly rugged energy landscape. 

The problem’s difficulty is reflected in the fact that optimal sequences have been verified only up to \(N = 66\)~\cite{packebusch2016labs}, underscoring its rapidly growing computational complexity. LABS can be formulated as the ground-state search of a long-range, four-local spin-glass Hamiltonian, representing a higher-order unconstrained binary optimization (HUBO) problem. This formulation has established LABS as a canonical testbed for benchmarking optimization algorithms across classical and quantum domains. A broad range of methods have been explored for LABS, including mixed-integer solvers such as CPLEX~\cite{cplex} and Gurobi~\cite{gurobi}, heuristic meta-optimizers such as the memetic tabu search (MTS)~\cite{labs_mts}, and quantum algorithms such as the quantum approximate optimization algorithm (QAOA)~\cite{evidence_qad} and recent approaches based on Pauli correlation encoding~\cite{sciorilli2025competitivenisqqubitefficientsolver}.

These approaches scale exponentially in computational effort, typically measured by the number of objective function evaluations. Among classical heuristics, the memetic tabu search (MTS) currently provides the best known performance, exhibiting a scaling of \(\mathcal{O}(1.34^{N})\) and reliably reaching optimal solutions up to \(N \le 64\)~\cite{BOSKOVIC2017262}. Despite this success, the exponential growth in cost ultimately limits the tractable sequence length, motivating the search for alternative paradigms. Quantum optimization methods have been investigated as potential routes to overcome this scaling barrier. The QAOA with twelve layers achieves a scaling of \(\mathcal{O}(1.46^{N})\) for \(28 \le N \le 40\), which can be reduced to \(\mathcal{O}(1.21^{N})\) when combined with quantum minimum finding~\cite{evidence_qad}. The bias-field digitized counterdiabatic quantum optimization (BF-DCQO) algorithm attains comparable scaling to twelve-layer QAOA while requiring up to six times fewer entangling gates for instances up to \(N = 30\)~\cite{koch2025quantumoptimizationbenchmarkinglibrary,cadavid2024biasfielddigitizedcounterdiabaticquantum,romero2024biasfielddigitizedcounterdiabaticquantum,chandarana2025runtime,iskay}. However, the practical reach of these quantum algorithms remains constrained by hardware limitations. Experimental realizations have so far demonstrated QAOA with a single layer up to \(N = 18\) on trapped-ion hardware~\cite{evidence_qad} and BF-DCQO up to \(N = 20\) on superconducting devices~\cite{koch2025quantumoptimizationbenchmarkinglibrary}. These restrictions highlight the need for hybrid quantum–classical schemes that can leverage the strengths of both domains to improve scaling within currently available computational resources.

In this work, we introduce a hybrid non-variational quantum-classical approach, the quantum-enhanced memetic tabu search (QE-MTS), to address the LABS problem. QE-MTS integrates DCQO~\cite{PhysRevApplied.22.054037} with MTS. The quantum stage generates low-energy candidate sequences that seed the initial MTS population, providing statistically biased starting points that guide the subsequent local search. 
The framework represents a concrete instance of hybrid sequential quantum computing~\cite{chandarana2025hybridsequentialquantumcomputing}, and quantum-enhanced optimization~\cite{cep}. This division of roles aligns with present hardware capabilities: quantum processors efficiently sample structured bitstrings, while classical algorithms perform the combinatorial refinement through recombination, mutation, and memory-guided search. To rigorously assess algorithm performance, we conduct a comprehensive scaling analysis by benchmarking the time-to-solution as a function of system size. Our results demonstrate that QE-MTS exhibits significantly more favorable scaling behavior than standalone MTS for $N \leq 37$, with the performance gap widening as system size increases.  

The remainder of this paper is organized as follows. Section~\ref{sec:labs} defines the LABS objective. Section~\ref{sec:qealgo} details the hybrid solver. Section~\ref{sec:results} presents empirical results for \(27 \le N \le 37\) together with the scaling analysis. Section~\ref{sec:conclusions} concludes and discusses future extensions. Additional implementation details are provided in the appendices.

\section{Low-Autocorrelation Binary Sequences}
\label{sec:labs}

The LABS problem aims to find binary sequences $\mathbf{s} = (s_1, \ldots, s_N) \in \{\pm 1\}^N$ that minimize the objective
\begin{equation}\label{eq:ising_labs}
E(\mathbf{s}) = \sum_{k=1}^{N-1} C_k^2, \quad \text{with} \quad C_k = \sum_{i=1}^{N-k} s_i s_{i+k},
\end{equation}
where $C_k$ denotes the $k$-th autocorrelation coefficient. The cost function penalizes large off-peak autocorrelations and favors sequences with sharp correlation profiles, which are valuable in applications such as radar and communication signal design.
Expanding Eq.~\eqref{eq:ising_labs} and mapping $s_i \mapsto \sigma_i^z$ yields a long-range Ising Hamiltonian representation that reads as
\begin{equation}\label{eq:labs_hamiltonian}
\begin{split}
H_f &= 2 \sum_{i=1}^{N-2} \sigma_i^z \sum_{k=1}^{\lfloor (N-i)/2 \rfloor} \sigma_{i+k}^z \\
&\quad + 4 \sum_{i=1}^{N-3} \sigma_i^z \sum_{t=1}^{\lfloor (N-i-1)/2 \rfloor} \sum_{k=t+1}^{N-i-t} 
\sigma_{i+t}^z \sigma_{i+k}^z \sigma_{i+k+t}^z,
\end{split}
\end{equation}
which includes both two- and four-body interactions, characteristic of higher-order spin-glass models~\cite{gardner1985spin}. The number of two- and four-body interaction terms increases quadratically and cubically with the system size, respectively
\begin{align}\label{eq:term_scaling_two}
n_{\text{two}}(N) &=  
\begin{dcases}
\frac{N}{2}\left(\frac{N}{2}-1\right), & \text{if }N\ \text{even},\\[2pt]
\left(\frac{N-1}{2}\right)^2, & \text{if }N\ \text{odd},
\end{dcases} \\ \label{eq:term_scaling_four}
n_{\text{four}}(N) &= 
\begin{dcases}
\frac{N}{12}\left(\frac{N}{2}-1\right)(2N-5), & \text{if }N\ \text{even},\\[2pt]
\frac{1}{24}(N-3)(N-1)(2N-1), & \text{if }N\ \text{odd}.
\end{dcases}
\end{align}
These expressions provide an estimate of the computational resources required for a given instance. A quantum algorithm requiring $G_{2q}$ and $G_{4q}$ two-qubit gates to represent the two- and four-body interactions present in~\eqlabel{eq:labs_hamiltonian} would involve $\mathcal{O}\!\left(G_{2q}N^2 + G_{4q}N^3\right)$ entangling operations. Given present coherence times and connectivity constraints, current quantum processors can reliably simulate systems of only up to approximately $N \lesssim 20$~\cite{koch2025quantumoptimizationbenchmarkinglibrary}.

The LABS problem exhibits both global bitflip and bit reversal symmetries
\[
(s_1, \ldots, s_N) \equiv (-s_1, \ldots, -s_N) \equiv (s_N, \ldots, s_1) \equiv (-s_N, \ldots, -s_1),
\]
which reduce the effective search space from \(2^N\) to \(2^{N-2}\) distinct configurations. The flip symmetry can be incorporated at the Hamiltonian level by fixing one spin, whereas the reverse symmetry is more difficult to encode. Although these symmetries can be exploited by both classical and quantum optimizers, they offer at most a constant-factor reduction in computational effort. Since this work focuses on scaling exponents, we do not apply symmetry-based instance reduction.

A distinctive feature of LABS is its high spectral degeneracy. The number of distinct energy levels grows at most polynomially as \(\mathcal{O}(N^3)\), while the configuration space increases exponentially as \(\mathcal{O}(2^{N-2})\). This degeneracy arises from the squared autocorrelation terms in Eq.~\eqref{eq:labs_hamiltonian}, which restrict energy values to multiples of four, with an upper bound of \(N(N-1)(2N-1)/6\), see Appendix~\ref{app:scaling}. The uniform anti-ferromagnetic couplings in the Hamiltonian induce frustration, further enhancing degeneracy.
\begin{figure}[!tb]
    \centering
    \includegraphics[width=\linewidth]{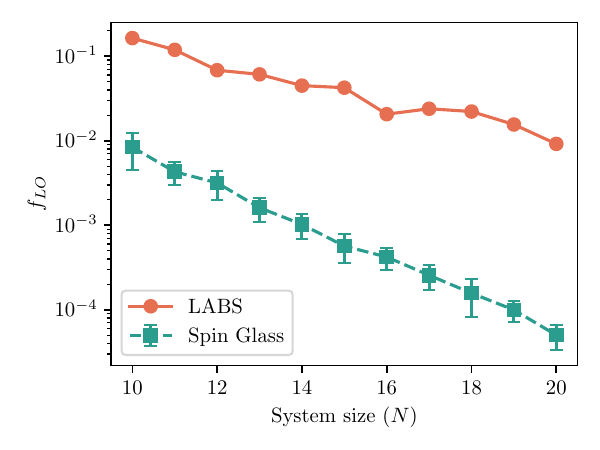}\vspace{-2mm}%
    \caption{Density of $1$-flip local minima $f_{\mathrm{LO}}$ for different $N$. For the spin-glass case, we report the results over $10$ independent instances. This quantifies the number of sequences in which no single flip reduces the energy, compared to the total number of configurations. Additionally, it supports why LABS is a more challenging problem than a common spin-glass benchmark.}
    \label{fig:local_minima_density}
\end{figure}%

These features produce a rugged energy landscape densely populated with local minima. We quantify this ruggedness through the density of single-flip local minima
\begin{equation}
    f_{\mathrm{LO}}(N) = 
    \frac{
        \left|\{\,s \in \{\pm1\}^N : E(s) \le E(s^{(i)})\ \forall i\,\}\right|
    }{2^N},
\end{equation}
where \(s^{(i)}\) is the configuration via flipping the \(i\)-th spin of \(s\).  This corresponds to the standard notion of one-spin-flip-stable (or metastable) states in mean-field spin glasses~\cite{bray1980metastable,aspelmeier2004complexity}, which provides a baseline for assessing ruggedness via the density of single-flip local minima of a given Hamiltonian. As shown in Fig.~\ref{fig:local_minima_density}, LABS exhibits a higher density of local minima than randomly generated Sherrington–Kirkpatrick (SK) spin-glass instances~\cite{sherrington1975solvable} with Gaussian couplings \(J_{ij} \sim \mathcal{N}(0, 1/N)\) and zero fields. This abundance of local traps creates problems for local search algorithms, where single-spin-flip dynamics converge to such metastable energies, and motivates our hybrid quantum-classical approach, combining nonlocal exploration via quantum sampling with classical refinement through metaheuristic search.

\section{Quantum-enhanced algorithm}
\label{sec:qealgo}
\subsection{Digitized counterdiabatic quantum optimization}
Adiabatic quantum optimization aims to prepare the ground state of the problem Hamiltonian $H_f$. An initial state is evolved within the time window $t\in[0,T]$ under the path $H_\text{ad}(\lambda)=(1-\lambda)H_i + \lambda H_f$. Here, $\lambda(t)$ toggles adiabatically the system, starting from an initial Hamiltonian $H_i$– whose corresponding ground state can be easily prepared– towards the problem Hamiltonian $H_f$, whose ground state encodes the solution of the optimization problem~\cite{lucas2014ising}. Relying on the adiabatic theorem, the ground state of $H_f$ is theoretically reached in the adiabatic limit $\dot{\lambda}(t)\to 0$. In the following, we use $H_i = \sum_i h_i^x \sigma^x_i$ as initial Hamiltonian, with $h_j^x$ the transverse fields acting on the $j$-th spin. In particular, we set $h_j^x=-1$ so that the ground state of $H_i$ is $\ket{\psi(0)}=\ket{+}^{\otimes N}$ with $\sigma^x\ket{\pm}=\pm\ket{\pm}$. 

To overcome the intrinsically slow adiabatic evolution, it is possible to introduce an auxiliary counterdiabatic (CD) driving contribution to accelerate the process while suppressing diabatic transitions~\cite{demirplak2003adiabatic,berry2009transitionless}. Several implementations have been proposed, with one approximating the adiabatic gauge potential by a nested-commutator series expansion~\cite{claeys2019floquet}, such that the system evolves under $H(\lambda) = H_\text{ad}(\lambda) + \dot{\lambda}A^{(l)}_\lambda$ with
\begin{equation}\label{eq:nc}
 A_\lambda^{(l)} = i\sum_{k=1}^l \alpha_k(\lambda)O_{2k-1}(\lambda),
\end{equation}
where $O_0(\lambda) = \partial_\lambda H_\text{ad}$ and $O_k(\lambda) = [H_\text{ad}, O_{k-1}(\lambda)]$. In the limit $l\to\infty$, this expansion converges to the exact gauge potential $A_\lambda$, which suppresses every possible transition. The coefficients $\alpha_k$ are obtained by minimizing the action $S_l=\text{Tr}{[G_l^2]}$ with $G_l=\partial_\lambda H_\text{ad} - i\big[H_\text{ad},A^{(l)}_\lambda\big]$ (see Appendix~\ref{app:nc}). In this work, we consider the first-order nested-commutator term ($l=1$), which takes the form
\begin{equation}\label{eq:1nc}
\begin{split}
-iO_{1}
&=4\sum_{i=1}^{N-2}\sum_{k=1}^{\left\lfloor\frac{N-i}{2}\right\rfloor}
\Big[h_i^x\sigma_i^y\sigma_{i+k}^z+h_{i+k}^x\sigma_{i+k}^y\sigma_i^z\Big] \\
&\quad+8\sum_{i=1}^{N-3}\sum_{t=1}^{\left\lfloor\frac{N-i-1}{2}\right\rfloor}
\sum_{k=t+1}^{\,N-i-t}
\sum_{p\in\{i,i+t,i+k,i+k+t\}}
h_p^x\sigma_p^y\!\!\!\prod_{\substack{q\in\{i,i+t,i+k,i+k+t\}\\ q\neq p}}\!\!\!\sigma_q^z.
\end{split}
\end{equation}

For a fast evolution, the adiabatic contribution $H_\text{ad}(\lambda)$ becomes negligible compared to $H_\text{cd}(\lambda)$ and therefore, can be omitted. This is known as the impulse regime and has been shown to be useful while designing quantum algorithms for optimization problems~\cite{PhysRevApplied.22.054037, chandarana_pf, PhysRevApplied.21.034033}, as it reduces the number of resources needed without compromising performance. Although the implementation of CD terms remains a challenge on current analog quantum platforms, their digitization has been leveraged in digital quantum computers~\cite{hegade2021shortcuts}. The resulting time-evolution operator can be represented as a quantum circuit via digitization and the first-order Trotter-Suzuki decomposition, i.e., $U(T,0) = \prod_{k=1}^{n_\text{trot}} \exp[\Delta t \, \alpha_1(k\Delta t) \, \dot{\lambda}(k\Delta t) \, O_1(k\Delta t)]$, where $n_\text{trot}$ is the number of Trotter steps and $\Delta t = T/n_\text{trot}$. In this work we take the limit $n_\text{trot}\to\infty$ to get the continuous time-evolution operator under $H(\lambda)$ in the impulse regime. After each circuit execution, all qubits are measured $n_\text{shots}$ times, returning binary bitstring $b_0\dots b_{N-1}$ with $b_i\in\{0,1\}$. Each measurement translates into an energy sample of $H_f$. We build the DCQO circuits using an efficient decomposition into two-qubit gates, see Appendix~\ref{app:nc}; and leverage CUDA-Q~\cite{cudaq} kernels to execute them using GPUs. Particularly, the simulation is done in the range $27\leq N\leq 37$ for $n_\text{shots}=10^5$ using an \textit{Amazon EC2 P6-b200.48xlarge} instance, see Appendix~\ref{app:gpus}.

\subsection{Quantum-enhanced memetic tabu search}\label{sec:mts}

Memetic tabu search (MTS) is a population-based metaheuristic that combines global exploration with local intensification (via tabu search)~\cite{Tabu}. The population maintains candidate sequences. Recombination and mutation explore new regions and then, tabu search performs a focused improvement of each offspring while preventing cycling through a short-term memory. In our quantum-enhanced MTS (QE-MTS) algorithm, DCQO supplies bitstrings that are used as the initial population, while the MTS baseline starts from a random one. Specifically, DCQO is run for a finite number of shots and the lowest bitstring is replicated $K$ times to form the initial population, see Algorithm~\ref{alg:ws-mts}. For the MTS baseline, all $K$ individuals are drawn randomly, which is the standard approach~\cite{labs_mts}. In both MTS and QE-MTS, we use $K=100$, $p_{\mathrm{comb}}=0.9$, $p_{\mathrm{mut}}=1/N$ and tournament size $2$. We run the MTS with a single thread to avoid over-estimation of the number of objective evaluations, and we stop the program as soon as the certified optimum is found. The time-to-solution (TTS) is measured the number of objective evaluations until the known optimal is for the first time. While this definition of TTS does not have units of time, it is possible to recover the exact runtimes from $\text{TTS} \times \tau$, where $\tau$ is the time to make one function evaluation.
\begin{algorithm}[!t]
\label{alg:ws-mts}
\small
\caption{Quantum-enhanced memetic tabu search}
\KwIn{Sequence length $N$; population size $K$ or initial population from a quantum sub-routine $P_Q$; recombination probability $p_{\text{comb}}$; maximum number of generations $G_\text{max}$; mutation rate $p_{\text{mut}}$; optional target energy $E_{\text{target}}$}
\KwOut{Best sequence $s^\star$ found}
\BlankLine
\textbf{Initialize}\;
\Indp
Use $P_\text{Q}$ as the initial population.
Set $s^\star$ to the best energy from the $P_\text{Q}$ individuals.\;
\Indm
\BlankLine
\While{$E(s^\star) > E_{\text{target}}$ and $G\leq G_\text{max}$}{
    \uIf{$\text{rand}(0,1) < p_{\text{comb}}$}{
        Select two parents $p_1, p_2$ (tournament)\;
        $\,c \leftarrow$ \textsc{Combine}$(p_1,p_2)$\tcp*{\footnotesize See Appendix~\ref{app:mts}}
    }\Else{
        $\,c \leftarrow$ a random individual from the population\;
    }
    \textsc{Mutate}$(c,p_{\text{mut}})$\tcp*{\footnotesize See Appendix~\ref{app:mts}}
    $\,c \leftarrow$ \textsc{TabuSearch}$(c)$\tcp*{\footnotesize See Appendix~\ref{app:mts}}
    \If{$E(c) < E(s^\star)$}{ $s^\star \leftarrow c$ }
    Replace a uniformly random individual in the population by $c$\;
    $G \leftarrow G+1$\;
}
\Return{$s^\star$}
\end{algorithm}

\section{Results}
\label{sec:results}

In this section, we run QE-MTS and perform a statistical analysis in order to predict the typical TTS of our approach. For each $N$ and each algorithmic variant (MTS and QE-MTS) we run multiple independent replicates. A replicate is a distinct initial parent pool: for QE-MTS it consists of a single quantum-enhanced initial population, whereas for MTS it is an independently sampled random pool. Within each replicate we run $100$ seeds and record the TTS.

\subsection{Scaling analysis}

Let $Y_{N,m,r,s}$ denote the TTS for sequence length $N$, method $m$, replicate $r$, and seed $s$. In our study, $27\leq N\leq 37$, $m\in\{\text{MTS},\text{QE-MTS}\}$, $r\leq 100$ and $s\leq 100$. Define the per-replicate summary as $\tilde Y_{N,m,r} = \operatorname{median}_{s}\{Y_{N,m,r,s}\}$, which yields a distribution of medians $\{\tilde Y_{N,m}\}_r$ for each pair $(N,m)$. We pick the medians to represent the TTS at a certain $N$ since they are stable against outliers and the question we aim to answer is: how long is one typically expected to wait until the optimal solution is found. Moreover, we characterize this distribution of medians via the quantiles $Q_{p}(N,m) = \operatorname{quantile}_{r}\!\big(\{\tilde Y_{N,m}\}_r;\,p\big)$, with $p\,\in\,\{0.10,0.50,0.90\}$. In particular, $Q_{0.50}(N,m)$ is the median of medians; $Q_{0.10}$ reflects a typical best-case (low TTS); and $Q_{0.90}$ includes typical worst-cases (high TTS). To quantify the scaling of $Q_p$ with respect to $N$, we fit the log-linear model
\begin{equation}
\label{eq:loglin}
\log Q_{p}(N,m) = \alpha_{m,p} + \beta_{m,p}\,N,
\end{equation}
which assumes $Q_{p}(N,m) \sim (\kappa_{m,p})^{N}$, where $\kappa_{m,p} = \exp(\beta_{m,p})$ is the exponential scaling base. 

To take into account the uncertainty coming from both replicates and seeds, we use a two-stage bootstrap. For each draw $b=1,\dots,B$ with $B=5000$: (i) resample replicates with replacement within each $(N,m)$; (ii) for each selected replicate, resample seeds with replacement and recompute $\tilde Y^{(b)}_{N,m,r}$; (iii) get $Q^{(b)}_{p}(N,m)$ across resampled replicates; (iv) fit Eq. (\ref{eq:loglin}) to obtain $\beta^{(b)}_{m,p}$ and the coefficients of determination $R^{2,(b)}_{m,p}$. This bootstrapping  mirrors how new data would be generated, including both between- and within-replicate variability, i.e. the variability coming from having different $r$ and different $s$, respectively. We report the results with $95\%$ confidence intervals for $\kappa$ and $R^2$ in~\tablabel{tab:scaling_cis}, where the median scaling base for QE-MTS is $\kappa\approx 1.24$ ($95\%$ CI $[1.23,1.25]$) versus $\kappa \approx 1.37$ ($95\%$ CI $[1.36,1.37]$) for MTS, indicating a shallower growth of typical TTS under QE-MTS. This suggests that scaling of MTS improves when it is quantumly enhanced. Besides, the scaling of the lower and higher quantiles, $Q_{0.10}$ and $Q_{0.90}$, also lie in a similar range of $\kappa$, indicating that the typical TTS is not expected to spread drastically as the system size grows. Nevertheless, quantiles lower than $Q_{0.10}$ may be affected by outliers, resulting in lower scaling but worse $R^2$ values. We also note that the obtained exponent bases may slightly change depending on the chosen range of $N$. For instance, including $N>37$ changes the scaling of MTS from $\mathcal{O}(1.37^N)$ to $\mathcal{O}(1.34^N)$. This small change indicates that the chosen range $(27\leq N\leq37)$ provides a reasonable reference.
\begin{table}[!tb]
\centering
\caption{Two-stage bootstrap $95\%$ CIs for $\kappa=\exp(\beta)$ and $R^2$ ($B=5000$ draws).}
\label{tab:scaling_cis}
\vspace{3pt}
\begin{ruledtabular}\begin{tabular}{llcc}
Method & Summary & $\kappa$ CI & $R^2$ CI \\
\midrule
QE-MTS & $Q_{0.50}$ & [1.23,\;1.25] & [0.86,\;0.89] \\
QE-MTS & $Q_{0.10}$ & [1.25,\;1.27] & [0.83,\;0.91] \\
QE-MTS & $Q_{0.90}$ & [1.24,\;1.25] & [0.90,\;0.92] \\
MTS    & $Q_{0.50}$ & [1.36,\;1.37] & [0.85,\;0.87] \\
MTS    & $Q_{0.10}$ & [1.34,\;1.36] & [0.85,\;0.88] \\
MTS    & $Q_{0.90}$ & [1.37,\;1.39] & [0.85,\;0.87] \\
\end{tabular}\end{ruledtabular}
\end{table}

\begin{figure}
\centering
\includegraphics[width=\linewidth]{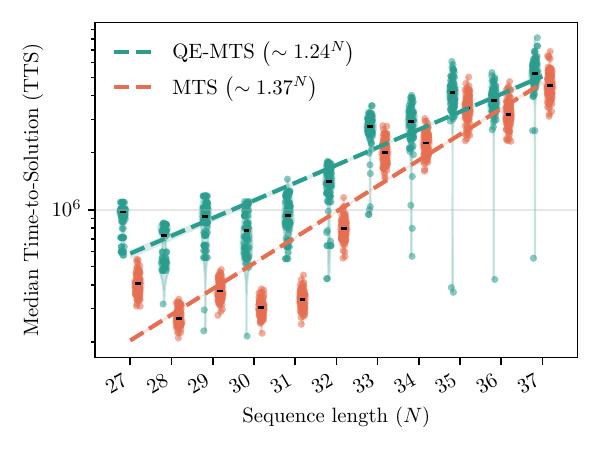}\vspace{-2mm}
\caption{Per-$N$ distributions of per-replicate medians $\tilde Y_{N,m,r}$ for QE-MTS (teal) and MTS (orange) on a log-scaled TTS axis. Each dot is a replicate median; dashed curves are log–linear fits to $Q_{0.50}(N,m)$ (median-of-medians), with fitted bases $\sim1.24^N$ (QE-MTS) and $\sim1.37^N$ (MTS). Lower positions indicate fewer function evaluations (faster).}
\label{fig:medians_distribution}
\end{figure}%

In Fig.~\ref{fig:medians_distribution}, we compare the typical TTS across different sequence lengths by plotting the distribution of per-replicate medians for each method. At smaller $N$, the MTS bulk data is notably lower than QE-MTS, indicating a smaller intercept, i.e., faster typical runs. Nevertheless, the trend of QE-MTS is a slower increase, implying a scaling advantage with respect to MTS. Moreover, we observe that in the region of $29\le N \le 37$ some of the QE-MTS medians lay at lower values with respect all their MTS relatives, which is specially remarkable for $33 \le N \le 37$. In fact, these outliers start to be captured by $Q_{0.04}$, where $\kappa$ drops down to $1.17$ at the cost of a lower quality fit with $R^2\approx 0.5$. In the studied sequence length range, QE-MTS remains slightly above MTS in typical TTS, even though the gap visually narrows. In Appendix~\ref{app:difference}, we quantify this gap by computing the log-ratio $\log_{10}(\mathrm{TTS}_{\mathrm{QE-MTS}})-\log_{10}(\mathrm{TTS}_{\mathrm{MTS}})$ across the system size; and in Appendix~\ref{app:reducing_tts} we present a simple alternative that sets the next steps to further lower the TTS such that the enhancement is two-fold, in the scaling as well as the intercept from~\eqlabel{eq:loglin}.

\subsection{Estimated crossover}
\label{sec:crossover}

To turn the qualitative pattern from Fig.~\ref{fig:medians_distribution} into a decision rule, we ask at what $N$ the scaling advantage becomes relevant, i.e., when QE-MTS is expected provide lower typical TTS values. For this, we define a conservative crossover $N_\times$ as the sequence length where the upper region of the QE-MTS data falls below the lower region of the MTS one. Concretely, from the fits (see~\eqlabel{eq:loglin}) of $Q_{0.95}(N,\text{QE-MTS})$ and $Q_{0.05}(N,\text{MTS})$. In each bootstrap draw $b$, $N_\times$ is computed 
\begin{equation}
\label{eq:crossover}
N_\times^{(b)} \;=\; \frac{\alpha_{\text{MTS},\,0.05}^{(b)}-\alpha_{\text{QE\text{-}MTS},\,0.95}^{(b)}}{\beta_{\text{QE\text{-}MTS},\,0.95}^{(b)}-\beta_{\text{MTS},\,0.05}^{(b)}}.
\end{equation}
The expected crossover is then estimated to occur at
\begin{equation}
\operatorname{median} N_\times \;=\; 46.598, \qquad 95\%\ \text{CI }[44.937,\;48.863].
\end{equation}
Therefore, under this conservative bulk criterion, QE-MTS becomes the safer choice at $N\gtrsim 47$. Note that this estimation does not include the cost of drawing the samples from the quantum computer, since we are interested solely in the effect of the initial quantum population. Nevertheless, in a practical scenario, it is important to include the time units of TTS, i.e., re-formulate it as $\text{TTS}=T_Q + T_C$. Here, $T_C = \tau_C \times \text{\# objective-evaluations}$, with $\tau_C$ the effective wall-clock time to evaluate once the objective function; and $T_Q = \tau_Q \times n_\text{shots}$, with $\tau_Q$ the effective wall-clock time to do one shot. While this shifts the crossover point in practice, the scaling improvement of QE-MTS remains.


\section{Conclusions}
\label{sec:conclusions}

We have presented a hybrid non-variational quantum-classical optimization algorithm, the quantum-enhanced memetic tabu search (QE-MTS), for the low-autocorrelation binary sequence problem. By initializing a classical memetic tabu search with samples generated through digitized counterdiabatic quantum optimization, QE-MTS exhibits an empirical scaling advantage over the best-known classical heuristic. In the range $27\leq N \leq37$, QE-MTS lowers the typical scaling from $\mathcal{O}(1.37^{N})$ to $\mathcal{O}(1.24^{N})$. The quantum stage achieves approximately a sixfold reduction in circuit depth compared to the QAOA, providing a more practical route toward realizing quantum advantage on early fault-tolerant hardware.

These findings demonstrate that combining problem-specific quantum samplers with high-performance classical metaheuristics can deliver measurable improvements in scaling for combinatorial optimization. More broadly, QE-MTS illustrates a class of quantum-enhanced optimization strategies in which quantum subroutines augment mature classical solvers, offering a realistic pathway toward achieving computational advantage in near-term quantum systems for real-scale and industrially relevant optimization problems.

Future work includes exploring adaptive DCQO variants to enhance the diversity of the initial population to decrease the typical time-to-solution while maintaining the observed scaling advantage. Finally, while our study evaluates a single quantum-classical pipeline, it does not discard the usage of other classical or quantum heuristics as the source of seeds.

\begin{acknowledgments}
We thank Bhargava Balaganchi and Anton Simen for valuable discussions. We thank Michael Falkenthal and Sebastian Wagner for their support with GPU access.
\end{acknowledgments}

\appendix

\section{Spectrum size of the LABS problem} \label{app:scaling}

To estimate the size of the spectrum, we start by showing that the objective function (\eqlabel{eq:ising_labs}) is upper bounded by $\mathcal{O}(N^3)$,

\begin{equation}
\begin{split}
    0\leq E(s) &= \sum_{k=1}^{N-1} \left(\sum_{i=1}^{N-k} s_i s_{i+k}\right)^2 \\
    &\leq \sum_{k=1}^{N-1} \left(\sum_{i=1}^{N-k} 1\right)^2 \\
    &= \frac{N(N-1)(2N-1)}{6} \\
    &= \mathcal{O}(N^3).
\end{split}
\end{equation}

Additionally, two distinct energy levels have a minimal separation of $\Delta E = E(s)-E(s')=4$, which follows from $s_is_{i+k} \pm s'_is'_{i+k} = 0,\pm2$ and

\begin{equation}
\Delta E = \sum_{k=1}^{N-1}\left(\sum_{i=1}^{N-k}(s_is_{i+k} + s'_is'_{i+k} )\right)\left(\sum_{i=1}^{N-k}(s_is_{i+k} - s'_is'_{i+k} )\right)
\end{equation}

This means that the energy levels are distributed in steps of $4$ while being upper bounded $\mathcal{O}(N^3)$. Therefore, the spectrum size also scales as $\mathcal{O}(N^3)$ in the worst case.

\section{Efficient construction of DCQO circuits}
\label{app:nc}

In this section, we describe how to build the DCQO circuits, which requires the computation of the first-order counterdiabatic coefficient, as well as the transpilation into quantum gates. The DCQO circuit is given by $U(T,0) = \prod_{k=1}^{n_\text{trot}} \exp[\theta(k\Delta t) \, O_1(k\Delta t)]$, where $\theta(t) = \Delta t\, \alpha(t) \, \dot{\lambda}(t)$ and 

\begin{figure}[!tb]
    \centering
    \includegraphics[width=0.25\linewidth]{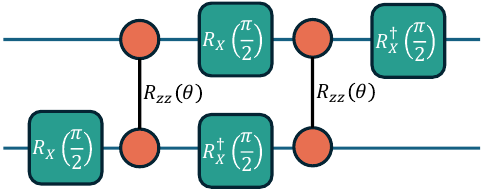}
    \caption{Decomposition of the block of two-qubit rotations $\text{R}_{YZ}(\theta)\text{R}_{ZY}(\theta)$, requiring $2$ entangling gates $\text{R}_{ZZ}$ and $4$ single-qubit gates.}
    \label{fig:two_body_circuit}
\end{figure}

\begin{widetext}

\begin{equation}
\begin{split}
   -i O_1 &= 4\sum_{i=1}^{N-2} h_i^x \sigma^y_i\sum_{k=1}^{\lfloor\frac{N-i}{2}\rfloor}\sigma^z_{i+k} + 4\sum_{i=1}^{N-2} \sigma^z_i\sum_{k=1}^{\lfloor\frac{N-i}{2}\rfloor}h_{i+k}^x \sigma^y_{i+k} \\
   &+ 8\sum_{i=1}^{N-3} h_i^x \sigma^y_i\sum_{t=1}^{\lfloor\frac{N-i-1}{2}\rfloor}\sum_{k=t+1}^{N-i-t}\sigma^z_{i+t}\sigma^z_{i+k}\sigma^z_{i+k+t} + 8\sum_{i=1}^{N-3} \sigma^z_i\sum_{t=1}^{\lfloor\frac{N-i-1}{2}\rfloor}\sum_{k=t+1}^{N-i-t}h_{i+t}^x \sigma^y_{i+t}\sigma^z_{i+k}\sigma^z_{i+k+t} \\
    &+8 \sum_{i=1}^{N-3} \sigma^z_i\sum_{t=1}^{\lfloor\frac{N-i-1}{2}\rfloor}\sum_{k=t+1}^{N-i-t}h_{i+k}^x \sigma^z_{i+t}\sigma^y_{i+k}\sigma^z_{i+k+t} + 8\sum_{i=1}^{N-3} \sigma^z_i\sum_{t=1}^{\lfloor\frac{N-i-1}{2}\rfloor}\sum_{k=t+1}^{N-i-t} h_{i+k+t}^x \sigma^z_{i+t}\sigma^z_{i+k}\sigma^y_{i+k+t}.
\end{split}
\end{equation}

After trotterization, we can rewrite $U(T,0)$ as 

\begin{equation}
\begin{split}
   U(T,0) &= \prod_{k=1}^{n_\text{trot}} \Bigg( \prod_{i=1}^{N-2}\prod_{k=1}^{\lfloor\frac{N-i}{2}\rfloor} \exp\left[4i \theta(k\Delta t) h_i^x \sigma^y_i \sigma^z_{i+k}\right] \exp\left[4i \theta(k\Delta t) h_{i+k}^x \sigma^z_i \sigma^y_{i+k}\right]\Bigg) \\
   &\qquad\times \Bigg(\prod_{i=1}^{N-3} \prod_{t=1}^{\lfloor\frac{N-i-1}{2}\rfloor}\prod_{k=t+1}^{N-i-t} \exp\left[8i \theta(k\Delta t) h_i^x \sigma^y_i \sigma^z_{i+t}\sigma^z_{i+k}\sigma^z_{i+k+t}\right] \exp\left[8i \theta(k\Delta t) h_{i+t}^x \sigma^z_i \sigma^y_{i+t}\sigma^z_{i+k}\sigma^z_{i+k+t}\right] \\
   &\hspace{3.2cm} \exp\left[8i \theta(k\Delta t) h_{i+k}^x \sigma^z_i \sigma^z_{i+t}\sigma^y_{i+k}\sigma^z_{i+k+t}\right] \exp\left[8i \theta(k\Delta t) h_{i+k+t}^x \sigma^z_i\sigma^z_{i+t}\sigma^z_{i+k}\sigma^y_{i+k+t}\right]\Bigg).
\end{split}
\end{equation}

Defining the generalized Pauli rotations as $\text{R}_{P}(x) = \exp\left(-ixP\right)$, it follows that

\begin{equation}
\begin{split}
U(0,T) &= \prod_{k=1}^{n_\text{trot}} \Bigg( \prod_{i=1}^{N-2}\prod_{k=1}^{\lfloor\frac{N-i}{2}\rfloor} \text{R}_{Y_iZ_{i+k}}\left(4 \theta(k\Delta t) h_i^x \right) \text{R}_{Z_iY_{i+k}}\left(4 \theta(k\Delta t) h_{i+k}^x\right)\Bigg) \\
&\quad\times \Bigg(\prod_{i=1}^{N-3} \prod_{t=1}^{\lfloor\frac{N-i-1}{2}\rfloor}\prod_{k=t+1}^{N-i-t} \text{R}_{Y_iZ_{i+t}Z_{i+k}Z_{i+k+t}}\left(8 \theta(k\Delta t) h_i^x \right) \text{R}_{Z_iY_{i+t}Z_{i+k}Z_{i+k+t}}\left(8 \theta(k\Delta t) h_{i+t}^x \right) \\
&\hspace{3.cm}\text{R}_{Z_iZ_{i+t}Y_{i+k}Z_{i+k+t}}\left(8 \theta(k\Delta t) h_{i+k}^x \right) \text{R}_{Z_iZ_{i+t}Z_{i+k}Y_{i+k+t}}\left(8 \theta(k\Delta t) h_{i+k+t}^x\right)\Bigg).
\end{split}
\end{equation}

The two-qubit block requires two entangling $\text{R}_{ZZ}$ gates and four single-qubit rotations to be implemented, see Fig.~\ref{fig:two_body_circuit}. The four-qubit block requires $10$ entangling $\text{R}_{ZZ}$ gates and $28$ single-qubit rotations to be implemented. Overall, the number of entangling gates to do a single trotter step of DCQO is equivalent to doing two layers of QAOA. To put these numbers in context, for $N=67$ QAOA($p=12$) circuits require $1.4$ million entangling gates whereas DCQO ones require $236.3$ thousand. 
\end{widetext}%
\begin{figure}[!tb]
    \centering
    \includegraphics[width=\linewidth]{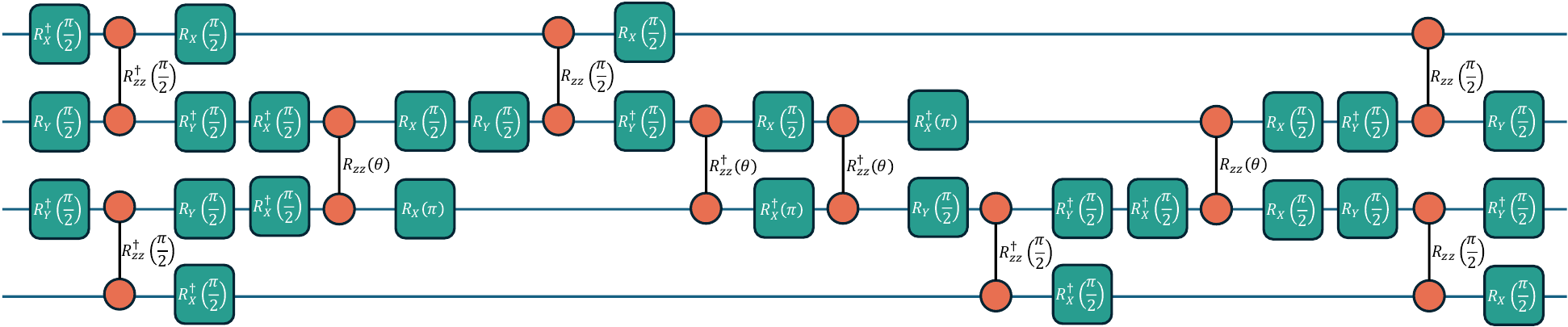}
    \caption{Decomposition of the block of four-qubit rotations $\text{R}_{YZZZ}(\theta)\text{R}_{ZYZZ}(\theta)\text{R}_{ZZYZ}(\theta)\text{R}_{ZZZY}(\theta)$, requiring $10$ entangling gates $\text{R}_{ZZ}$ and $28$ single-qubit gates.}
    \label{fig:four_body_circuit}
\end{figure}%

Now, the numerical computation of the first-order counterdiabatic coefficient $\alpha(t)$ can become computationally expensive even for moderately large system sizes due to the cubic scaling of the Hamiltonian terms, as shown in~\eqlabel{eq:term_scaling_four}. In the following lines, we describe how it was analytically computed to make the circuit preparation more efficient.

Recall the total Hamiltonian $H(\lambda)=H_\text{ad}(\lambda) + i\dot{\lambda}\alpha_1(\lambda)O_1(\lambda)$ from Sec.~\ref{sec:qealgo}. The coefficient $\alpha_1(\lambda)$ is obtained as $\alpha_1(\lambda) = -\Gamma_1(\lambda)/\Gamma_2(\lambda)$ with $\Gamma_k(\lambda) = \text{tr}\big[O_k^\dagger(\lambda)O_k(\lambda)\big]$. In particular, $\Gamma_1(\lambda)$ reads as
\begin{equation}
\begin{split}
\Gamma_1(\lambda)=\sum_{\alpha\in\{2,4\}} c_\alpha \sum_{A\in G_\alpha}\mathcal S_x(A),
\qquad c_2=16,\;c_4=64.
\end{split}
\end{equation}
where $\mathcal S_x(A)=\sum_{p\in A}(h_p^x)^2$ and $G_{2,4}$ the sets including all the two- and four-body interactions present in~\eqlabel{eq:labs_hamiltonian}, respectively. Following the same procedure for computing $\Gamma_2(\lambda)$,
\begin{widetext}
\begin{equation}
\begin{aligned}
\Gamma_2(\lambda)
&=-256\Bigg[
64\,\lambda^2\,\mathcal I_{44}
+\sum_{A\in G_2}\Big(\lambda^2\,\mathcal S_x(A)
+(1-\lambda)^2\big(\mathcal S_{bx}(A)+4\,\mathcal P_x(A)-\mathcal S_x(A)^2\big)\Big)\\
&\qquad\qquad
+4\sum_{A\in G_4}\Big(4\,\lambda^2\,\mathcal S_x(A)
+(1-\lambda)^2\big(\mathcal S_{bx}(A)+4\,\mathcal P_x(A)-\mathcal S_x(A)^2\big)\Big)
+4\,\lambda^2\big(4\,\mathcal I_{24}+\mathcal I_{22}\big)
\Bigg],
\end{aligned}
\end{equation}

where $\mathcal S_{bx}(A)=\sum_{p\in A}(h_p^x h_p^b)^2$, $\mathcal P_x(A)=\sum_{p<q\in A}(h_p^x h_q^x)^2$ and $\mathcal I_{\alpha\beta}(\bm{a},\bm{b})\coloneqq\big\{a_p \,|\, a_p=b_q, \forall (p,q)\in[1,\alpha]\times[1,\beta]\big\}$, with $\bm{a}\coloneqq (a_0,\dots,a_\alpha)\in G_\alpha$, $\bm{b}\coloneqq(b_0,\dots,b_\beta)\in G_\beta$ and $\alpha,\beta\in\{2,4\}$.
\end{widetext}

\section{Distance between MTS and QE-MTS}\label{app:difference}

\begin{figure}
\centering
\includegraphics[width=\linewidth]{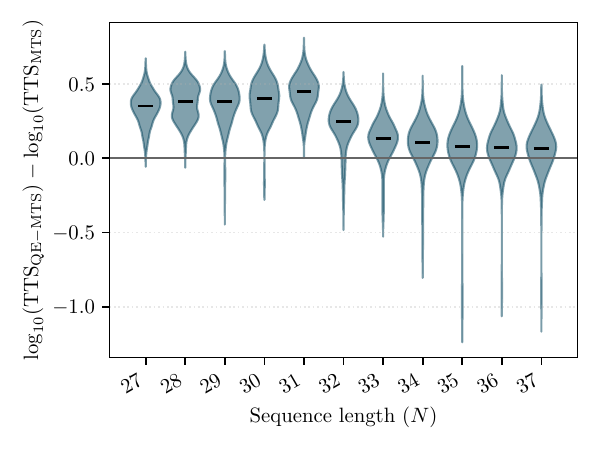}\vspace{-2mm}
\caption{Per-$N$ distributions of $\log_{10}(\mathrm{TTS}_{\mathrm{QE-MTS}})-\log_{10}(\mathrm{TTS}_{\mathrm{MTS}})$. Negative values indicate QE-MTS is faster (lower TTS) on a log scale. Distributions are obtained via a two-stage bootstrapping (replicates and seeds, $5000$ draws per $N$).}
\label{fig:diff_distribution}
\end{figure}

In Sec.~\ref{sec:results} we observed that the QE-MTS data approaches the MTS one, implying that the gap between the data points decreases until an eventual crossover. Here, we quantify this gap by computing the log-ratio $\log_{10}(\mathrm{TTS}_{\mathrm{QE-MTS}})-\log_{10}(\mathrm{TTS}_{\mathrm{MTS}})$ across the studied system sizes. Negative values of this ratio favor QE-MTS and since this metric is logarithmic, vertical differences have a direct multiplicative meaning: a value of $D$ implies $\mathrm{TTS}_{\mathrm{QE-MTS}}/\mathrm{TTS}_{\mathrm{MTS}}= 10^{D}$. In Fig.~\ref{fig:diff_distribution} we show the distribution of ratios across different sequence lengths. We observe that at small $N$, the distributions cluster slightly above. However, as $N$ increases, the difference shifts systematically to negative values, indicating a growing advantage for QE-MTS. This is consistent with the smaller scaling coefficients $\kappa$ in Table~\ref{tab:scaling_cis} and QE-MTS approaching MTS in Fig.~\ref{fig:medians_distribution}. In conclusion, although plain MTS typically performs better at small sizes, the typical advantage shifts toward QE-MTS as $N$ grows. 

\begin{figure}
    \centering
    \includegraphics[width=\linewidth]{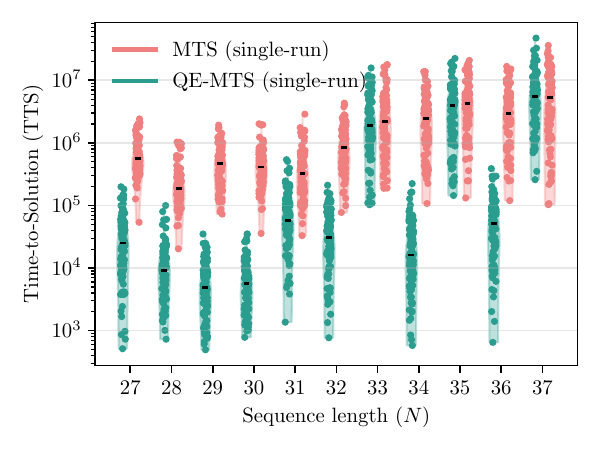}\vspace{-2mm}%
    \caption{Single-run comparison of time-to-solution (TTS) with sequence length $N$ using an alternative QE-MTS, where the initial population is taken from multiple DCQO runs (teal), and MTS (orange).}
    \label{fig:single_run_comparison}
\end{figure}

\section{Lowering time-to-solution}
\label{app:reducing_tts}

One specific observation from Fig.~\ref{fig:medians_distribution} is that the TTS values for QE-MTS lie above those of MTS for system sizes in the range $27 \leq N \leq 37$. In other words, the TTS values are effectively higher in this regime, even though the slope is lower. As discussed in the main text, the crossover between QE-MTS and MTS is expected to occur around a sequence length of $N \approx 47$. This apparent discrepancy arises from the specific warm-starting procedure used to initialize MTS with quantum-enhanced solutions. Alternative initialization strategies could, in principle, lower the TTS values and bring them below the MTS reference line.

In this section, we describe one such alternative setup of QE-MTS that achieves lower TTS across multiple system sizes. This motivates future work to identify an optimal configuration where the TTS remains low while preserving the scaling advantage. In this setup, we construct the initial population in Algorithm~\ref{alg:ws-mts} using the lowest-energy bitstrings obtained from multiple independent DCQO runs. This approach makes the initialization more robust against shot noise. Figure~\ref{fig:single_run_comparison} shows the performance of a single run of this QE-MTS variant compared to a typical MTS run, demonstrating that high-quality initial populations can indeed be obtained via DCQO. For most sequence lengths, the best-performing QE-MTS seeds outperform the best MTS seeds by up to two orders of magnitude. For three system sizes ($N = 33, 35, 37$), however, no enhancement was observed. Nevertheless, these isolated cases do not outweigh the consistent improvements across other system sizes and do not preclude further enhancements for larger $N > 37$.

\section{GPU hardware specifications}
\label{app:gpus}
\begin{figure}[h]
\centering
\includegraphics[width=1\linewidth]{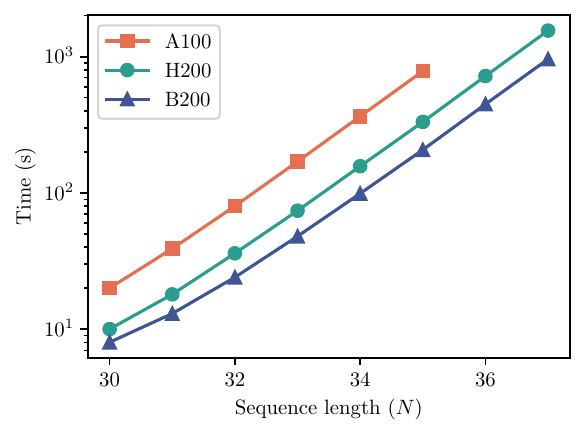}
\caption{Runtime comparison for NVIDIA A100, H200 and B200 GPUs for the simulation of the DCQO circuits through CUDA-Q~\cite{cudaq}.}
\label{fig:h200a100}
\end{figure}

We tested \textit{AWS p4d.24xlarge} (A100), \textit{p5en.48xlarge} (H200) and \textit{P6-b200.48xlarge} (B200) instances. The B200 and H200 GPUs offer higher memory, enabling $N=37$ in our DCQO simulations, see Fig.~\ref{fig:h200a100} and Table~\ref{tab:instances_aws}. 

\begin{table}[!tb]
\centering
\caption{EC2 instance specifications.}
\label{tab:instances_aws}
\begin{ruledtabular}\begin{tabular}{cccccc}
Instance & GPUs & vCPUs & RAM & GPU memory & Max $N$ \\ \midrule
A100     & 8    & 96    & 1\,TB & 320\,GB    & 35 \\
H200     & 8    & 192   & 2\,TB & 1128\,GB   & 37 \\
B200     & 8    & 192   & 2\,TB & 1440\,GB   & 37 \\
\end{tabular}\end{ruledtabular}
\end{table}

\begin{algorithm}[!tb]
\label{alg:Tabu}
\small
\caption{\textsc{TabuSearch}}
\KwIn{starting sequence $s_0$}
\KwOut{locally improved sequence $\tilde{s}$}
\BlankLine
$\,\tilde{s} \leftarrow s_0$; $p \leftarrow s_0$\;
Initialize \textit{TabuList}$[1\!:\!N]\gets 0$\;
$\,\textit{M} \leftarrow \text{random\_int}(0,N) + \lfloor N/2 \rfloor$ \tcp*{$\in \left[\tfrac{N}{2},\tfrac{3N}{2}\right]$}
\For{$t=1$ \KwTo $\textit{M}$}{
    Choose index $i^\star$ with minimum energy among $\{\,i \mid \textit{TabuList}[i] < t\,\}$\;
    $\,p \leftarrow \textsc{Flip}(p,i^\star)$\;
    $\theta_\text{min},\theta_\text{max} \leftarrow \text{M}/10,\, \text{M}/50$
    $\textit{TabuList}[i^\star] \leftarrow t + \text{random\_int}(\theta_{\min},\theta_{\max})$\;
    \If{$E(p) < E(\tilde{s})$}{ $\tilde{s} \leftarrow p$ }
}
\Return{$\tilde{s}$}
\end{algorithm}%
\begin{algorithm}[!tb]
\label{alg:ops}
\small
\caption{\textsc{Combine} and \textsc{Mutate}}
\SetKwProg{Fn}{Function}{:}{}
\Fn{\textsc{Combine}$(p_1,p_2)$}{
    Choose cut point $k\in\{1,\dots,N-1\}$ uniformly\;
    \Return{$p_1[1\!:\!k]\;\Vert\;p_2[k+1\!:\!N]$}\;
}
\vspace{0.5em}
\Fn{\textsc{Mutate}$(s,p_{\text{mut}})$}{
    \For{$i=1$ \KwTo $N$}{
        \If{$\text{rand}(0,1)<p_{\text{mut}}$}{ $s \leftarrow \textsc{Flip}(s,i)$ }
    }
\Return{$s$}
}
\end{algorithm}

\section{Tabu search}
\label{app:mts}

Tabu search is a metaheuristic for combinatorial optimization that extends greedy local search with a short‐term memory to avoid cycling and to encourage exploration. Recently visited states are declared tabu for a limited tenure so that the search can traverse non-improving regions without immediately undoing itself; an aspiration criterion lifts tabu status when a move yields a new global best \cite{Glover1989-Tabu-Search-Part-I, Glover1990-Tabu-Search-Part-II}. In our instantiation, Algorithm~\ref{alg:Tabu} (\textsc{TabuSearch}) iteratively inspects the one–flip neighborhood, selects the best admissible move under a bounded randomized tenure with aspiration, updates the incumbent when an improvement occurs, and runs for a randomized budget. Algorithm~\ref{alg:ops} provides the simple \textsc{Combine} (single–point crossover to seed restarts) and \textsc{Mutate} (independent flips with probability $p_{\text{mut}}$ to perturb incumbents) operators that generate starting points for the local improver.

\bibliography{reference.bib}
\end{document}